\documentclass[review]{elsarticle}
\usepackage{listings}
\usepackage[svgnames]{xcolor}
\usepackage{graphicx,amsmath,amssymb,float}
\usepackage{latexsym,natbib,url,hyperref}
\usepackage{wrapfig,lipsum,booktabs}
\usepackage{caption,subcaption}
\usepackage{multirow,enumitem}

\usepackage{multicol}

\DeclareMathOperator*{\argmin}{\arg\!\min}

\usepackage{bbm,float}
\usepackage{tikz}

\usepackage{geometry}
\date{}

\begin{document}

\begin{frontmatter}
\title{How do mobility restrictions and social distancing during COVID-19 affect the crude oil price?}

%\title{How mobility restrictions and social distancing during COVID-19 affect oil price}

\author[1,2]{Asim K. Dey}
\author[3]{Kumer P. Das}

\address[1]{Princeton University,Princeton, NJ, USA}
\address[2]{The University of Texas at Dallas, Dallas, TX, USA}
\address[3]{University of Louisiana at Lafayette,Lafayette, LA,USA}

\begin{abstract}
We develop an \textit{air mobility index} and use the newly developed Apple's \textit{driving trend index} to evaluate the impact of COVID-19 on the crude oil price. We use quantile regression and stationary and non-stationary extreme value models to study the impact. We find that both the \textit{air mobility index} and \textit{driving trend index} significantly influence lower and upper quantiles as well as the median of the WTI crude oil price. The extreme value model suggests that an event like COVID-19 may push oil prices to a negative territory again as the air mobility decreases drastically during such pandemics.

\end{abstract}

\begin{keyword}
	 \ Apple Mobility, \ Air Mobility,  \ Crude Oil,  \ Quantile Regression, \ Extreme Value Theory 
	%\MSC[2010] 00-01\sep  99-00
\end{keyword}

\end{frontmatter}

\section{Introduction}
Understanding the real-time fuel demand is very important for many reasons. Energy merchants want to get a trading edge from all possible resources such as thermal images from cameras on pipelines, satellite data tracking worldwide oil tankers, and so on. Another way of quantifying the real-time demand is by tracking the transportation sector which accounts for the largest share of U.S. petroleum consumption. In 2019, the transportation sector accounts for 68\% of all petroleum consumption in the U.S. Moreover, finished motor gasoline accounts for about 45\% of total U.S. petroleum consumption (U.S. Energy Information Administration). Human mobility trends introduced by Apple Inc. in mid-April can also be used to understand on-the-spot gasoline consumption data as it captures user activity in searching for directions on smartphones. In particular, it is of great interest to use such mobility data during and after the COVID-19 lockdowns where the fuel demand fluctuates vigorously~\citep{Bildirici2020AnalyzingCO,Jefferson2020ACF,Shiqi2020}.

Apple has introduced the tool to support the impactful work happening around the globe to mitigate the spread of COVID-19~\citep{appleMob}. 
Using aggregated data collected from Apple maps, this trend is generated by counting the number of requests made for directions. 

To quantify mobility restrictions and social distancing we use Apple mobility trends, in particular \textit{driving trend} in US~\citep{appleMob}. 
The other important component of fuel consumption is jet fuel consumption which has been decreased drastically due to COVID-19. Although an average of 24,900 commercial passenger flights departed U.S. airports each day in January 2020, by July 2020, flight volume had declined to 13,700 per day--51\% of the July 2019 level~\citep{EIAB2020}. 
We introduce a new \textit{air mobility index} based on temporal airline network to evaluate the dynamics of daily U.S. flight volumes. A low score of the two indices indicates that people are following the government's guidelines on mobility restrictions more and practicing more social distancing. A high score of the two indices implies lower mobility restrictions and less social distancing. The objective of this study is to understand the impact of the Apple \textit{driving trend index} and the newly developed  \textit{air mobility index} on the crude oil price.

In this paper, first, we investigate the effect of driving trend and air mobility on the different quantiles of the crude oil price based on the quantile regression model. Second, we evaluate the tail of the crude oil price using an extreme value model. Generally, extreme value models focus on the upper tail of the distribution. However, we
study the lower tail of the crude oil price and analyze different determinants of extreme lower crude oil price based on non-stationary extreme value model as discussed in Dey et al. (2020)~\citep{Dey2020}.

The rest of the paper is organized as follows. In Section~\ref{sec:QMS} we introduce different measures of mobility restrictions and social distancing. In Section~\ref{sec:Data} we describe the data. We discuss the two methods, namely, quantile regression and extreme value theory, used in the study in Section~\ref{sec:Meth}. We report the findings and describe the results in Section~\ref{sec:result}. Finally, we conclude in Section~\ref{sec:concl}.

%The jet fuel is one of the main final product of crude oil. COVID-19 has substantially reduce passenger flight volumes. Consequently, jet fuel consumption has been declined significantly. 

%===============================================================================

%\subsection{Crude oil price}
%\section{Crude Oil Price and Mobility Trend}

\section{Quantification of mobility restrictions and social distancing}
\label{sec:QMS}
To control the spread of COVID-19 different governments have been implementing a variety of mobility restrictions and social distancing measures~\citep{Charoenwong2020,FANG2020104272,Pan2020,Setti2020}. 
To quantify the impact of such control measures on human mobility behavior we utilize the two following mobility indices.

\subsection{Apple's driving  trend index}

The Apple driving trend represents a relative driving volume of Apple users compared to a baseline volume on January 13, 2020. 
In our study we consider weekly average of the Apple driving trend between January 13, 2020, and August 25, 2020 as a measure of mobility. We name this measure as \textit{Apple driving trend index} ($ADI$). 
We normalize $ADI$ in week $t$ by adjusting the average $ADI$ and corresponding standard deviation as
\begin{eqnarray}\label{eq:Dr}
%M_t= {{Dr_t-  \mu_{Dr}} \over {\sigma_{Dr}}},
H_t= {{ADI_t-  \mu_{ADI}} \over {\sigma_{ADI}}},
\end{eqnarray}
where, $\mu_{ADI}$ and $\sigma_{ADI}$ are the mean and standard deviation of $ADI$ between January 13, 2020, and August 25, 2020.

%\newpage
\subsection{Air mobility index}

In our study, we consider daily flight volumes in the U.S. as a measure of air mobility. To evaluate daily flight volumes, we introduce a complex network analysis. We define a graph $G=(V, E)$ as a model for a airline network, with node set $V$ and set of edges $E \subset V \times V$ such that $(u,v) \in E$ represents an edge from node $u$ to $v$, $u,v=1,2, \cdots, n$.
Here, the nodes represent the U.S. airports and if there is a direct flight between two nodes (i.e., airports) they are connected by an edge. Here $|V|$ is the number of nodes and $|E|$ is the number of edges in the network. The elements of the $n\times n$-symmetric \textit{adjacency matrix}, $A$, of $G$ can be written as
\begin{equation}
A_{ij}=\begin{cases} 1, & \mbox{if}\; (i,j)\in E \\
0, & \mbox{otherwise}.
\end{cases}
\end{equation}
 We assume that $G$ is \textit{undirected} i.e., for all $e_{uv} \in E$,  $e_{uv}\equiv e_{vu}$.
 The degree, $d_u$, of a node $u$ is the number of edges incident to $u$ i.e., for u,v $\in$ V and e $\in$ E, $d_u = \sum_{u \neq v} e_{u,v}$.
We define $f_d$ to be the fraction of nodes $u \in V$ with degree $d_u = d$. The collection $\{f_d\}$, $d \le 0$ is called the degree distribution of $G$~\citep{Kolaczyk:Csardi:2014}.

\textit{Temporal Network} is an extension of network analysis which appears in many domains of knowledge such as finance~\citep{Battiston2010,Zhao2018}, epidemiology~\citep{Demirel2017,Dey2020VID19,Enright2018,Valdano2015}.  
A temporal network is a network structure that changes in time, which can be represented with a time indexed graph $G^t=(V(t),E(t))$, where, $V(t)$ is the set of nodes in the network at time $t$, $E(t) \subset V(t) \times V(t)$ is a set of edges in the network at time $t$. Here, $t$ is either discrete or continuous. To quantify the impact of COVID-19 on flight volumes which necessarily determines the jet fuel demand, we construct two temporal airline network ($G^t$) for two different time scale.

First, we develop a weekly airline network ($G^t_w$) in each week Sunday day ($t$) between January 13, 2020 and August 25, 2020: $\mathbb{G}_w = \{{G}^{1}_w, \ldots, {G}^{\mathcal{T}}_w\}$, where $\mathcal{T}=32$. We consider 
the airline network on each Sunday is an average representation of that week. The number of nodes ($|V^w_t|$), number of edges ($|E^w_t|$), and degree distribution are evaluated in each $G^t_w$.

\begin{figure*}[!ht]
     \centering
     \begin{subfigure}[b]{0.55\textwidth}
         \centering
         \vspace*{-15mm}
    \includegraphics[width=1.0\textwidth]{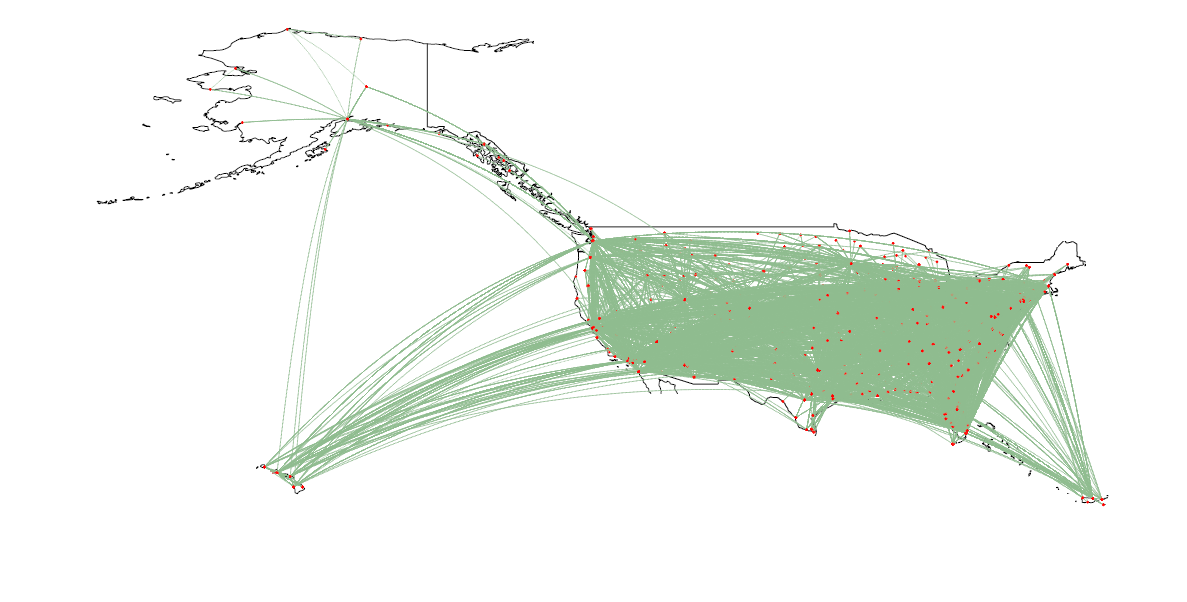}
         \caption{US airline network on Feb 2, 2020.}
         \label{fig:USnetA}
     \end{subfigure}
     \hfill
     \begin{subfigure}[b]{0.44\textwidth}
         \centering
         \includegraphics[width=1.0\textwidth]{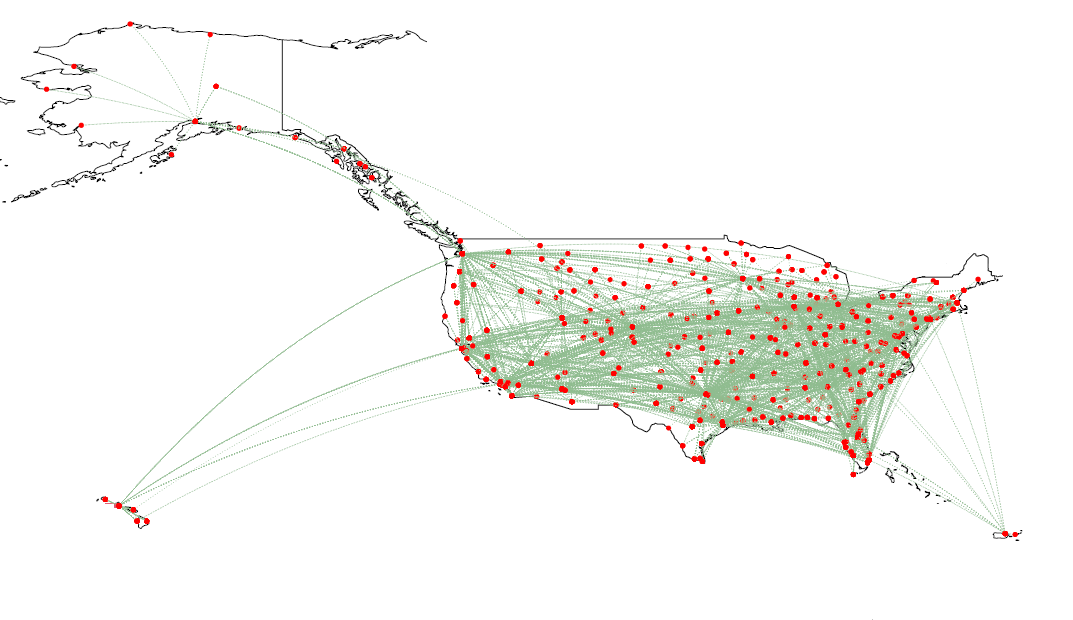}
         \caption{US airline network on May 3, 2020.}
         \label{fig:USnetB}
     \end{subfigure}
        \caption{Impact of COVID-19 on US airlines. Red points represent nodes (airports) and green lines represent edges. (a) Shows normal US airline network with 340 nodes and 18,805 edges.  
        (b) ) Represents reduced US airline network with 319 nodes and 4,980 edges.  }
     	\label{fig:Airline_Networks}
\end{figure*}

Figure~\ref{fig:Airline_Networks} illustrates the U.S. flight volumes on Feb 2, 2020, and on May 3, 2020. Table~\ref{table:QR_Tab1} shows how does the degree distribution of the US Airline network change (decrease) due to the COVID-19 pandemic. For example, the mean number of departed and landed flights in US airport on Feb 2, 2020, is 111, which reduces to 31 on May 3, 2020. The maximum number of departed and landed flights in a US airport on Feb 2, 2020, was 1,748, whereas,  the maximum number of departed and landed flights in a US airport on May 3, 2020, was only 683.

\begin{table*}[!ht]
%\small
\caption{Degree distribution of U.S. Airline networks on Feb 2, 2020 and on May 3, 2020.}
	\label{table:QR_Tab1}
	\centering
	\begin{tabular}{c | *{7}{c} r}  
	\hline
      Network  	       &   Min.& 1st Qu. & Median &   Mean & 3rd Qu.  &  Max. \\
      
	\hline
     %  &  &  &  &    &   &  \\
    $G^{02-02-2020}$ &   2 &    6 &   19 &  111   &  73 & 1748  \\ 
		\hline
	%	       &  &  &  &    &   &  \\
    $G^{05-03-2020}$   & 1 &   3  &   7 &   31 &   20 &   683  \\

	 \hline	

\end{tabular}
\end{table*}

%\begin{figure}[!ht]
%\centering
%  \includegraphics[width=0.70\textwidth]{Covid19CrudeOilPrice/Figures/Network_degree_dist.png}
%	\caption{Network degree dist.}
%	\label{fig:Degree_dist}
%\end{figure}

Second, we construct a monthly airline network ($G^t_m$) on $15^{th}$ of each month ($t$) between January 2000 and August 2020: $\mathbb{G}_m = \{{G}^{1}_m, \ldots, {G}^{\mathcal{T}}_m\}$, where $\mathcal{T}=248$. We consider 
the airline network on $15^{th}$ of each month is an average representation of that month. Similarly as $\mathbb{G}_w$, we evaluate the number of nodes ($|V^m_t|$), number of edges ($|E^m_t|$), and degree distribution in each $G^t_m$.

In our study, we consider the number of edges in an airline network at time $t$ as a measure of air mobility. 
We name the metrics 
%and is referred to 
as \textit{air mobility index} (AMI). Therefore we define the weekly and monthly air mobility index as $AMI^w_t = |E^w_t|$ and $AMI^m_t = |E^m_t|$, respectively. 
Similar to Eq.~\ref{eq:Dr}, we normalize $AMI^w$  and $AMI^m$ as 

\begin{eqnarray}\label{eq:GT}
K^w_t= {{AMI^w_t-  \mu_{AMI^w}} \over{\sigma_{AMI^w}}}   \quad\mathrm{and}\quad  K^m_t= {{AMI^m_t-  \mu_{AMI^m}} \over{\sigma_{AMI^m}}},
\end{eqnarray}

where $\mu_{AMI^w}$, $\mu_{AMI^m}$ and $\sigma_{AMI^w}$,$\sigma_{AMI^m}$ are the mean and standard deviation of $AMI^w$ and  $AMI^m$ respectively.

\subsection{COVID-19 and Mobility trend}

We study the impact of COVID-19 variables ($\textbf{C}$), e.g., weekly U.S. new cases, weekly U.S. new deaths on U.S. mobility dynamics and social distancing. We standardized each COVID-19 variable as

\begin{eqnarray}\label{eq:CVt}
O_t= {{C_t-  \mu_C} \over{\sigma_{C}}},
\end{eqnarray}
where, $C_t$ is a COVID-19 variable (weekly US new cases, weekly US new deaths) in week $t$,  $\mu_C$ and $\sigma_{C}$ are the mean and standard deviation of the corresponding variable between January 13, 2020, and August 25, 2020.

\begin{figure*}[!ht]
\centering
  \includegraphics[width=0.70\textwidth]{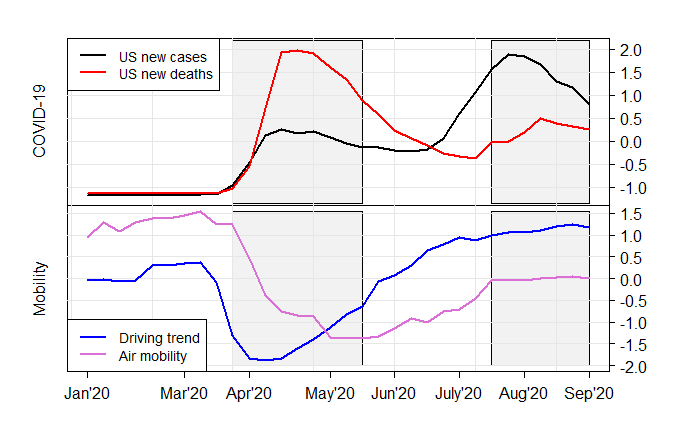}
	\caption{Time plots of normalize weekly average of COVID-19 variables (new cases and new deaths in the US) and normalize weekly average mobility metrics from January 2020 to September 2020.}
	\label{fig:COVID19_Index_timeplot}
\end{figure*}

Figure ~\ref{fig:COVID19_Index_timeplot} shows the time plots of normalize weekly average of COVID-19 new cases and deaths in 2020, along with the two mobility indices, i.e., weekly driving trend and weekly air mobility in 2020. Notice that in the first gray region, between late March and middle of May, when the COVID-19 new cases and deaths jump, the driving trend and air mobility collapse. However, in the second gray region, between the middle of July and the end of August, even though there are surges in new cases and deaths, driving trend and air mobility increase gradually. That is, initially a sudden increase in COVID-19 new cases and deaths negatively influence the driving trend and air mobility, but after a certain period, high COVID-19 new cases and deaths do not cause lower driving and decrease air mobility.

%The line graphs extend the understanding of the COVID-19 impact on medium-term fuel demand as the pandemic evolves.

%========================================

\section{Data}
\label{sec:Data} 

The WTI crude oil price (\$ per Barrel) is obtained from the U.S. Energy Information Administration~\citep{EIA_Data}. 
We compute weekly average of WTI crude oil price, $P'$, between January 13, 2020, and August 25, 2020.
We normalize weekly WTI crude oil price $P'_t$ in week $t$ as
\begin{eqnarray}\label{eq:WTI}
P_t= {{P'_t-  \mu_{P'}} \over {\sigma_{P'}}},
\end{eqnarray}
where, $\mu_{P'}$ and $\sigma_{{P'}}$ are the mean and standard deviation of $P'$ between January 13, 2020, and August 25, 2020.

Apple mobility data are obtained from Apple mobility trends reports~\citep{appleMob}. Air mobility index is based on two types of data: airline On-Time Performance Data and airport coordinate data. The flight data are obtained from the Bureau of Transportation Statistics (BTS)~\citep{BTS2020}. 
And, U.S. airport coordinate data are obtained from a geospatial data management platform named Koordinates~\citep{Koordinates2020}. 
We obtain U.S. COVID-19 data from \textit{Our World in Data}~\citep{ourworld2020}.

%============================================================================
%============================================================================

\section{Methodology}
\label{sec:Meth}

We evaluate the impact of driving trends and air mobility on WTI crude oil prices. First, we apply the quantile regression methodology to evaluate the effects of driving trend and air mobility volume on WTI crude oil price.
Second, we quantify the left tail or extreme lower quantiles of the WTI crude oil price associated with the COVID-19 pandemic with nonstationary extreme value models.

\subsection{Quantile Regression}

Quantile regression is a generalization of linear regression.
Whereas the classical least squares regression estimates the conditional mean of the response variable, quantile regression estimates the conditional median and other quantiles of the response variable. That is, quantile regression constructs a set of regression lines for different quantiles of the conditional distribution of the dependent variable.
Unlike least squares regression, quantile regression does not depend on the particular parametric assumption for the dependent variable and it also avoids constant variance assumption~\citep{Fasiolo2020,Koenker2001,Koenker2002,Nicolai2006}.

We can model a $\tau^{th}$ quantile of a response variable $y$ given a particular value of the predictor variable, $\mathbf{X}=\mathbf{x}$, as 
\begin{eqnarray}\label{eq:QR}
Q_Y (\tau|\mathbf{x}) = \beta_0(\tau)  + \beta_1(\tau) x_1 + \beta_2(\tau) x_2,
\end{eqnarray}
where $\tau\in$ (0, 1), $Y$ stands for weekly WTI crude oil price ($P$), $X_{1}$ is driving trend ($H$), and $X_{2}$ is air mobility ($K^w$)~\citep{koenker_2005}.

The parameters can be estimated by solving the equation
\begin{eqnarray}\label{eq:LF1}\nonumber
\hat{V} (\tau) = \argmin_{b} \sum\limits_{i=1}^n \rho_\tau (y_i - x_i^T b) 
\end{eqnarray}
where, 
\begin{equation}\nonumber
\rho_\tau (z) =\begin{cases} \tau z, & \mbox{if}\; z \geq 0 \\
(\tau - 1) z, & \mbox{if}\;  z < 0.
\end{cases}
\end{equation}

Koenker and Machado (1999) define a goodness of fit of the quantile regression model by comparing its error sum of squares with error sum of squares of a restricted form of the model. In our study, we define a restricted model which only include an ``intercept''
as $Q_Y (\tau|\mathbf{x}) = \beta_0(\tau)$ with 
$\tilde{V} (\tau) = \argmin_{b_0} \sum\limits_{i=1}^n \rho_\tau (y_i - x_i^T b_0)$. We  compute the Pseudo $R^2$ of the model defined in Equation ~\ref{eq:QR} as
\begin{eqnarray}\label{eq:R1}
R^2 (\tau) = 1 - {\hat{V} (\tau) \over \tilde{V} (\tau)}.
\end{eqnarray}

\subsection{Extreme Value Model}

Extreme value models evaluate the tail of a distribution and they have a wide range of applications in climate and atmospheric science to industrial risks, geosciences, finance, economics, and insurance~\citep{castrocamilo2018,Castillo2005,Chen2019,Das2016,Dey2016,Brian2019,reich2012}.

For a sequence of independent random variables, $Y_1, Y_2,...,Y_n$, with a common distribution function $G$, let $M_n=\max\left\lbrace Y_1, Y_2,...,Y_n\right\rbrace$ be the maximum of the process over $n$ time units (i.e., \textit{block}) of the observations. For $\{a_n >0\}$ and $\{b_n\}$ then $Pr\left\lbrace \left(M_n - b_n \right)/a_n \leq z\right\rbrace \rightarrow G(z)$ as $n \rightarrow \infty$, where $G$ is a non-degenerate function.
According to the Fisher and Tippett extremal theorem~\citep{fisher_tippett_1928}, 
$G$ belongs to the family of Generalized Extreme Value (GEV) distribution. The distribution function of a non-stationary GEV model can be written as
\begin{eqnarray}\label{eq:Q3} 
F\left(z;\mu(t),\sigma(t),\xi(t)\right)  &=& \exp\left\lbrace-\Big[1+\xi(t) \Big(\frac{z-\mu(t)}{\sigma (t)}\Big)^{-\frac{1}{\xi(t)}}\Big]\right\rbrace,
\end{eqnarray}

The GEV parameters $\mu(t)$, $\sigma(t)$, and $\xi(t)$ are functions of time or/and other covariates. The $t$ denotes the time (e.g., months, year, etc.) over which the maximum is chosen.  A more detail about the nonstationary GEV model and its' parameters can be found in~\citep{Coles2001,Dey2020}. 
 The $r$-year return level, i.e., the level expected to be exceeded once every $r$ year, can be defined as 
\begin{eqnarray}\label{retun_2}
z_r=\mu(t) + {\sigma(t)\over \xi(t)} \left[ {\left( -\log \left(1- r^{-1}\right)\right) ^{-\xi(t)}}-1\right].
\end{eqnarray}

The probability of an extreme event ($z$) can be obtained as
\begin{eqnarray}\label{prob_1}
P\left(Z>z\right)= 1 - \exp \left\lbrace -\left(  1 - \xi(t)  \left( \frac{\mu(t)-z}{\sigma(t)}\right) ^{-\frac{1}{\xi(t) }}\right)  \right\rbrace.
\end{eqnarray}

If $\mu(t)=\mu$, $\sigma(t)=\sigma$ and $\xi(t)= \xi$, the non-stationary GEV model becomes stationary GEV model. We can estimate the GEV model parameters using maximum likelihood estimation (MLE). The concept of return period and return level is used to quantify the likelihood of extreme events. 

%========================================================

In our study, we are interested in modeling the lower tail of the WTI crude oil price, $P$, to evaluate the impact of social distancing and mobility restrictions (i.e, $ADI$ and $AMI$) on it. That is, we focus on the minima of WTI crude oil price over blocks rather than its maxima. The same GEV model settings can be applied here based on the relation,
%$M'_n=\min\left\lbrace Y_1, Y_2,...,Y_n\right\rbrace =-\max\left\lbrace -Y_1, -Y_2,...,-Y_n\right\rbrace$.

\begin{eqnarray}\label{EVT:Min}
M'_n=\min\left\lbrace Y_1, Y_2,...,Y_n\right\rbrace =-\max\left\lbrace -Y_1, -Y_2,...,-Y_n\right\rbrace.
\end{eqnarray}

By rearranging, we find
\begin{eqnarray}\label{EVT:Min}
-M'_n=-\min\left\lbrace Y_1, Y_2,...,Y_n\right\rbrace = \max\left\lbrace -Y_1, -Y_2,...,-Y_n\right\rbrace.
\end{eqnarray}
%\newpage 

Therefore, to evaluate the lower tail of WTI crude oil price we use the GEV model on negative block minima of WTI crude oil price~\citep{JSSv072i08}.

\section{Results}\label{sec:result}
%\subsection{Data and Variables}\label{sec:Data}

We examine the effect of mobility restrictions and social distancing due to COVID-19 on WTI crude oil price. We quantify mobility restrictions and social distancing based on Apple's driving trend Index and a newly proposed Air mobility index. 

\begin{figure*}[!ht]
\centering
\includegraphics[width=0.75\textwidth]{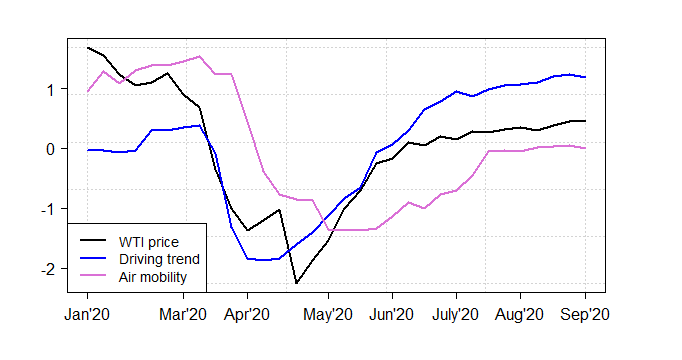}
	\caption{Time plots of normalize weekly average crude oil price (\$ per Barrel) and normalize weekly average mobility metrices from January 2020 to August 2020.}
	\label{fig:WTI_Index_timeplot}
\end{figure*}

Figure~\ref{fig:WTI_Index_timeplot} depicts the dynamics of WTI crude oil price,  Apple's driving index, and Air mobility index from January 2020 to September 2020. It is clear from the figure that the divining trend and air mobility significantly decreased in March and April of 2020, primarily due to the surge of COVID-19 and related stay-at-home orders at different states in the U.S. during that period. These confinement measures slumped the demand for oil and oil products, which results in a collapse in WTI crude oil price. Although the driving trend increased significantly after restrictions eased, air mobility did not increase much primarily because of people's tendency to avoid public transportation. WTI crude oil price gradually increased after lockdown and other restrictions were relaxed.

%--------------------------- DL ----------------------\\

We now evaluate the effect of driving and air mobility tend on WTI crude oil price. We develop a quantile regression model for each of the quantiles $\tau=(0.01,0.05,0.1,0.5,0.9,0.95,0.99)$ based on Eq.~\ref{eq:Q3}. We are particularly interested in the lower and upper quantile of the WTI crude oil price.
Table~\ref{table:QR_Tab2} summarizes the outputs of the quantile regression models.

We find that both driving trend and air mobility significantly affects lower and upper quantiles as well as the median of the WTI crude oil price. The coefficients of driving trend and air mobility are all positive. Therefore, a decrease (increase) in driving trend or air mobility leads to a decrease (increase) in
$\tau^{th}$ quantiles of the WTI crude oil price, where $\tau \in (0.01,0.05,0.1,0.5,0.9,0.95,0.99)$. Notice that the coefficient of driving trend tends to gradually decrease from the lower quantiles to upper quantiles. On the other hand, the coefficient of air mobility tends to gradually increase from the lower quantiles to the upper quantiles. That is, the driving trend has a greater influence on the lower quantiles of the WTI crude oil price. On the other hand, air mobility has a higher influence on the upper quantiles of the WTI crude oil price.

\begin{figure*}[!ht]
\centering
  \includegraphics[width=0.70\textwidth]{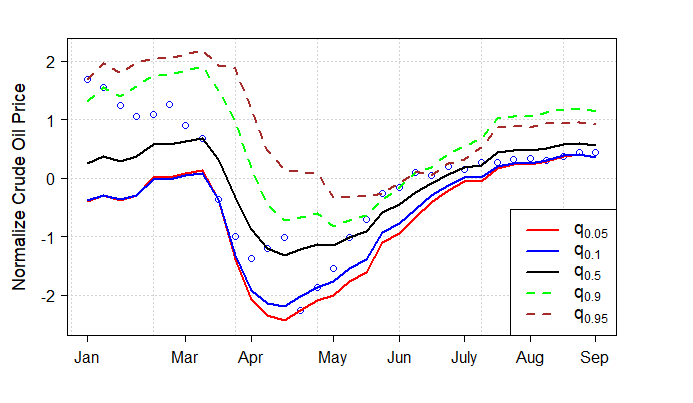}
	\caption{Normalize WTI crude oil prices regression models for different quantile levels.}
	\label{fig:QR2}
\end{figure*}

The values of $R^2$ are higher in lower quantiles. That is,  driving trend and air mobility able to explain the variability of lower quantiles of WTI crude oil price better compared to its upper quantiles. In our study, we focus on how does driving trend and air mobility shape the collapse of WTI crude oil price during the COVID-19 pandemic.
The $R^2$ values reveal that the driving trend and air mobility
greatly influence the extreme decrease of WTI crude oil price (lower quantiles) during the pandemic.
Figure~\ref{fig:QR2} depicts the regression lines of WTI crude oil price for different quantiles.

%\section{Quanlile Regression}

\begin{table*}[!ht]
%\small
\caption{Estimates of quantile regression models for crude oil price with standard errors. $^{***}~p<0.01$, ~$^{**}~p<0.05$,~$^{*}~p<0.1$.}
	\label{table:QR_Tab2}
	\centering
	\begin{tabular}{c | *{7}{c} r}  
	\hline
		       &     &   &   & Quantile ($\tau$) &  &  &  \\
       	       & 0.01     & 0.05   &  0.10 & 0.50 & 0.90  & 0.95 & 0.99 \\
	\hline
    Constant ($\beta_0$)  & -0.649$^{***}$ &  -0.649$^{***}$ & -0.568$^{***}$ & -0.071 & 0.405$^{***}$ & 0.585$^{***}$ &  0.585$^{***}$ \\ 
                          &  (0.041)       & (0.097)        &  (0.114)        & (0.185) & (0.102)       & (0.144)       & (0.074) \\ 
		\hline
    Driving trend ($\beta_1$)        & 0.846$^{***}$  &  0.846$^{***}$ &  0.786$^{***}$ &  0.527$^{***}$ & 0.559$^{***}$ &  0.273$^{*}$ &  0.273$^{*}$ \\ 
                                     &  (0.034)      & (0.081)         &  (0.090)       & (0.178)         & (0.151)       & (0.229)     &  (0.169) \\ 
       	\hline
    Air mobility ($\beta_2$)    &  0.298$^{***}$ &  0.297$^{**}$ &  0.229$^{*}$ &  0.358$^{**}$ &  0.514$^{***}$ & 0.617$^{***}$ &   0.617$^{***}$  \\ 
                                &  (0.077)       & (0.145)        & (0.170)      & (0.199) &  (0.085)            &  (0.088)      & (0.039) \\ 
      	\hline

     Pseudo $R^2$  & 0.738 &   0.722 &  0.687 & 0.536 & 0.540 &  0.523 & 0.526 \\  

	 \hline	

    		 \hline	
\end{tabular}
\end{table*}

%====================================================================================

%\newpage
%\subsection{Extreme Value Model-- Results}

\begin{figure*}[!ht]
\centering
  \includegraphics[width=0.70\textwidth]{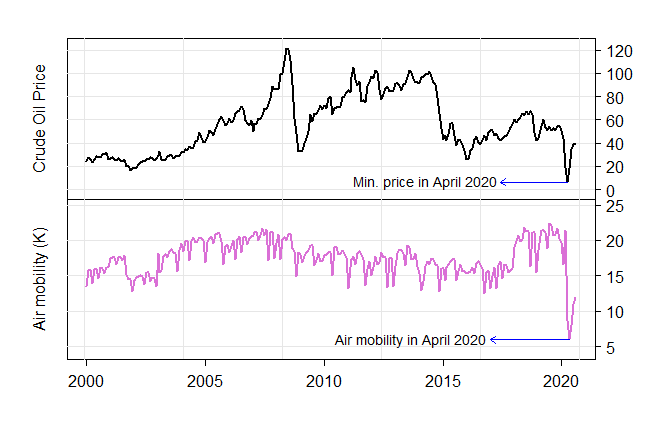}
	\caption{Monthly minimum crude oil price and monthly air mobility from January 2000 to August 2020.}
	\label{fig:GR_SNP_timeplot}
\end{figure*}

Now we turn our analysis to evaluate the crash of WTI crude oil price resulting from the COVID-19 pandemic based on the extreme value model. We study monthly minimum crude oil price (Y) for the last twenty years from 2000 to 2020 and model left tail, i.e., extreme lower quantile of the WTI crude oil price.
However, due to the unavailability of Apple's driving trend index, we only use monthly air mobility $K^m$ as a covariate in the extreme value modeling. Figure~\ref{fig:GR_SNP_timeplot} shows the time plot of monthly minimum crude oil price and monthly air mobility (on the $15^{th}$ of every month) in the U.S. from January 2000 to August 2020.

We experiment a set of of non-stationary extreme value models, $Y'_t \sim GEV(\mu(t), \sigma(t),  \xi(t))$, where $Y'_t=-Y_t$, $Y_t$ is the monthly minimum crude oil price in month $t$.
We assess the different combinations of time trends and covariate (air mobility) effect on the monthly minimum crude oil price. We select two best models based on  (smaller) Akiake Information Criterion (AIC)~\citep{Akaike1998} 
and Bayesian information criterion (BIC)~\citep{schwarz1978} values.
Table~\ref{table:GEVmodels} describes the selected models along with a baseline stationary model.
Model 1 allows a linear time trend in the location parameter. In Model 2 location parameter depends linearly on time trend and air mobility. There is a time trend in the log-scale parameter in both models.

\begin{table*}[!ht]
	\caption{Different 	stationarity/non-stationary models with corresponding AIC and BIC.}
	\label{table:GEVmodels}	
	%	\small
	\centering
	\begin{tabular}{c|c|*{6}{c}r}  
		
		Model  &   Model descriptions  &  AIC/BIC   \\
		\hline

		Model 0 &  $Y'_t \sim GEV(\mu, \sigma, \xi)$ & $AIC = 2285.212$\\
		(Stationary) &    & $BIC = 2295.751$  \\

%		\hline	
%	 &  $Y_t \sim GEV(\mu (t), \sigma (t), \xi)$  &  $AIC=2275.821 $ \\
%	Model 1	&    $\mu(t)=\beta_0 + \beta_1 t + \beta_2 t^2$  &  $BIC =  2289.875$  \\

		\hline	
	 &  $Y'_t \sim GEV(\mu (t), \sigma (t), \xi)$  &  $AIC=2164.935$ \\
	Model 1	&    $\mu(t)=\beta_0 + \beta_1 t$   &  $BIC = 2182.503$  \\
		&    $log~\sigma(t)= \gamma_0 + \gamma_1 t$  &  \\
		
				\hline
		 &   $Y'_t \sim GEV(\mu (t), \sigma(t), \xi)$ & $AIC = 2129.58$ \\
		Model 2 &    $\mu(t)=\beta_0 + \beta_1 t + \delta K^m$  & $BIC = 2150.66$  \\
		&    $ log~\sigma(t)= \gamma_0 + \gamma_1 t$ & \\
		\hline
		
		\hline
	\end{tabular}
\end{table*}
Table~\ref{table:EVT_estmates} shows the estimated parameters of Model~1 and Model~2. In Model~2 the covariate air mobility has a negative effect on monthly minimum WTI crude oil price, where the estimated value of  $\delta$ is -5.190, with a standard error of 0.974.
Therefore, one unit increase in air mobility moves monthly minimum WTI crude oil price in opposite direction (decrease) by 5.190 units when covariate ($t$) held constant. That is, an increase in air mobility results in an increase in overall WTI crude oil price.

Figure~\ref{fig:Diag_EVT} in Appendix represents the goodness of fit plots of the two models. We find that Quantile-Quantile (Q-Q) plots for both Model~1 and Model~2 approximately $45$-degree line. The observed versus fitted density plot for both models also yield that both Model~1 and Model~2 capture the variability of the monthly minimum WTI crude oil price.

\begin{table*} [!ht]
	\caption{Estimated parameters of nonstationary GEV models described in   Table~\ref{table:GEVmodels}, standard errors are in parenthesis.}
	\label{table:EVT_estmates}	
	\begin{center}
		\begin{tabular}{l*{6}{c}r} \hline
			
		& 	$\hat{\beta_0}$ &  $\hat{\beta_1}$  & $\hat{\delta}$  & $\hat{\gamma}_0$ & $\hat{\gamma}_1$ &  $\hat{\xi}$   \\
			\hline
		Model~1	& -22.11 &   -0.387 &  & 1.974 & 0.010 &  -0.606 \\
			& (0.651) & (0.014) & & (0.056)  & ($<$ 0.001) &   (0.0443) \\
			\hline

	Model~2	& -27.965 & -0.298 & -5.190 & 2.035 & 0.008 & -0.631\\
			& (0.909) & (0.015) & (0.974) & (0.053) & ($<$ 0.001) & (0.049)  \\
			\hline 
			
		\end{tabular}
	\end{center}
\end{table*}

%-----------------------------------------------------------

We compute the return level, i.e., the extreme lower quantile of WTI crude oil price which is expected to
fall below on average once every $r$ year using Eq.~\ref{retun_2}.
The return levels $z_r$ based on Model~1 depend on time $t$. That is, for certain values of $t$ we can compute return levels of crude oil price for different return periods.
The return levels based on Model~2 depend on time $t$ as well as on monthly air mobility.
Table~\ref{return_level2} represents two scenarios of the covariates and their corresponding 10, 20, 50, and 100 years return levels.

\begin{table*} [!ht]
	\caption{Estimated return levels for the GEV distribution fitted to the monthly minimum crude oil price.  $t=193$ (Jan-16), $t=244$ ( Apr-20).}
	\label{return_level2}	
	%\small
	\begin{center}
		\begin{tabular}{l*{6}{c}r} \hline
			
			& Covariates    & 10-yr return  & 20-yr return  & 50-yr return  & 100-yr return    \\
			& values &   level & level &  level  &  level  \\
			
					\hline
		                 %	& $t=108$ & 44.254  & 39.399  &  34.070 &  33.002\\
	            	Model~1	& $t=193$ (Jan 2016)    & 42.042 & 35.387 &  30.139 & 27.754 \\
			                & $t=244$  (Apr 2020)  & 27.826  &  17.049 &  8.5499 & 4.687\\

			\hline
              % 	& $t=108$, $D=23.57$ & 42.038 & 37.816  & 33.485  & 32.685 \\
	   Model~2	& $t=193$, $K^m=15308$ (Jan 2016)   & 35.253 & 29.911 & 25.779 & 23.939 \\
			    & $t=244$, $K^m=8643$  (Apr 2020)  & 12.687 & 4.517 &  -1.801 & -4.615\\
			    
			\hline 
		\end{tabular}
	\end{center}
\end{table*}

\begin{figure*}[!ht]
	\centering			
		\includegraphics[width=0.65\textwidth]{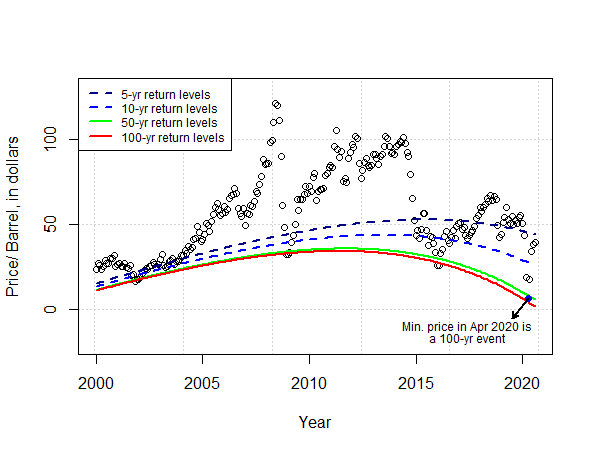}
	\caption{Return levels of minimum crude oil prices for different return periods, based on Model~1.}
	\label{RTL:M1}
\end{figure*}

\begin{figure*}[!ht]
	\centering			
		\includegraphics[width=0.70\textwidth]{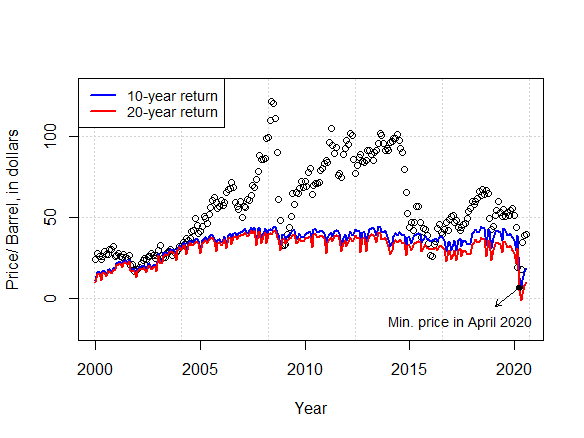}
	\caption{Return levels of minimum crude oil prices for different return periods, based on Model~2.}
	\label{RTL:M2}
\end{figure*}

Based on Model~1, for $t$ corresponds to April 2020, the estimate of the crude oil price that we expect to fall below once every
10, 20, 50, and 100 years are 27.826, 17.049, 8.5499, and 4.687, respectively.
Notice that the minimum crude oil price in April 2020 was 6.65 and the $100$-year return level is 4.687. Therefore, based on Model~1, we can say that the minimum oil price in April 2020 is a $100$-year event, which is also clear in Figure~\ref{RTL:M1}.
However, when we include monthly air mobility as a covariate in the model (Model~2), we find that return levels fall dramatically.
For the values of $t$ and $N$ in April 2020, the 10 and 20 years return levels are 12.687 and 4.517, respectively. Therefore, lower air mobility ( Model~2) makes the minimum oil price in April 2020 a $20$-year event. That is, lower air mobility increases the risk of occurring a very low oil price.
Figure~\ref{RTL:M2} depicts how a very low oil demand (i.e., oil demand in April 2020) causes frequent lower oil prices.

We have also found that Model~2 predicts 50-yr return and 100-yr return levels as -1.801 and -4.615, respectively. Therefore, there is a risk of occurring a negative oil price -1.801 in every 50 years, and a negative oil price -4.615 in every 100 years, if there is very low air mobility, i.e., $K^m=8,643$ (with a time trend $t=244$).

\section{Conclusion}\label{sec:concl}
To the best of the author's knowledge, the proposed \textit{air mobility index} and Apple's \textit{driving trend index} have not been used to study the impact of COVID-19 on crude oil prices. We use the concept of \textit {temporal network} to develop the \textit{air mobility index}. We use quantile regression to evaluate the effects of driving trend and air mobility volume on WTI crude oil price. Based on the results obtained from quantile regression, we claim that driving trend has a higher influence on lower quantiles of crude oil prices whereas air mobility has a higher influence on upper quantiles of crude oil prices. We also quantify the tail or extreme quantiles of the WTI crude oil price associated with the COVID-19 pandemic by stationary and non-stationary GEV models. We conclude that with COVID-19, an already volatile oil market has reached a flashpoint. Although oil prices have recovered significantly in the past couple of months,  it is unlikely that there will be the same buoyancy in prices as witnessed following the previous global economic recessions such as the Great Recession in 2008.

%---------------------Appendix----------------
\newpage
\section{Appendix}
\label{Appnd}

\begin{figure*}[!ht]
\centering
\begin{subfigure}{.40\textwidth}
  \centering
  \includegraphics[width=.99\linewidth]{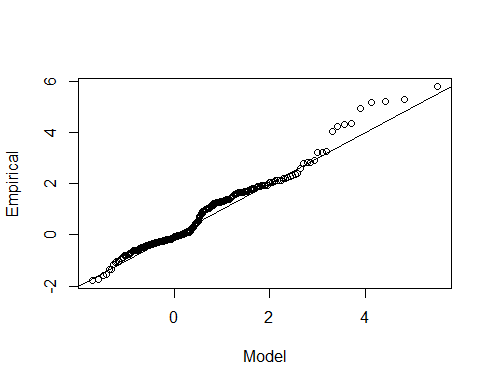}
  \caption{Q -- Q plot for Model~1.}
  \label{fig:qq}
\end{subfigure}%
\begin{subfigure}{.40\textwidth}
  \centering
  \includegraphics[width=.99\linewidth]{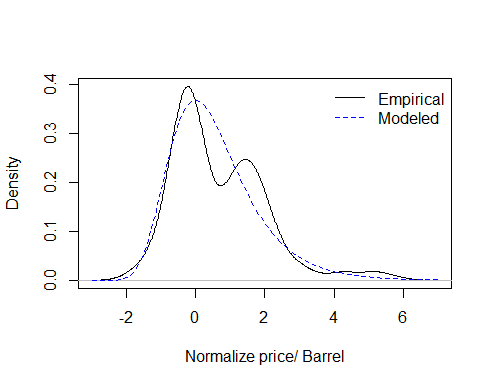}
  \caption{Density plot, Model~1.}
  \label{fig:density}
\end{subfigure}

\begin{subfigure}{.40\textwidth}
  \centering
  \includegraphics[width=.99\linewidth]{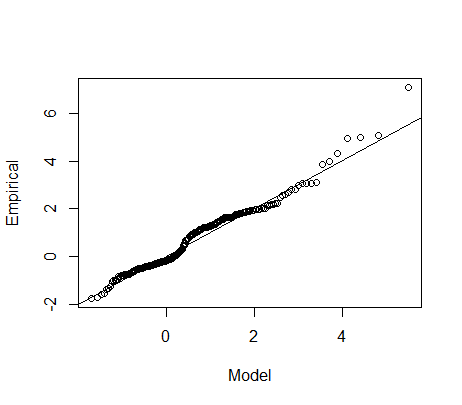}
  \caption{Q -- Q plot for Model~2.}
  \label{fig:qq2}
\end{subfigure}%
\begin{subfigure}{.40\textwidth}
  \centering
  \includegraphics[width=.99\linewidth]{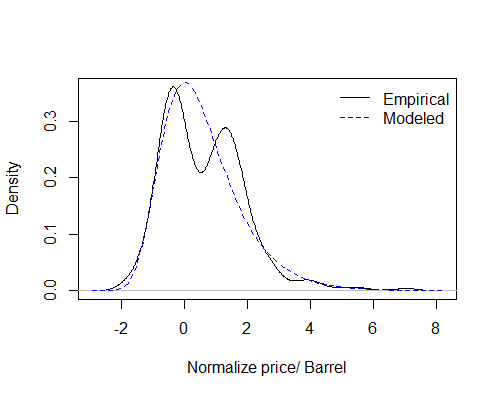}
  \caption{Density plot, Model~2.}
  \label{fig:density2}
\end{subfigure}
\caption{Diagnostic plots for Model~1 and Model~2.}
\label{fig:Diag_EVT}
\end{figure*}

%=============================================================================================
\newpage
\clearpage
\bibliographystyle{elsarticle-harv}\biboptions{authoryear,comma}
\bibliography{Covid_V2}

\end{document}